\documentstyle[12pt]{article}
\begin{document}

\begin{center}
{\large \bf Gravitational Lensing Effects of Fermion-Fermion Stars:\\
                I. Strong Field Case }\\
\vspace{8mm}
Ke-Jian Jin$^{1,3}$, Yuan-Zhong Zhang$^{2,1^{*},4,}$
       and  Zong-Hong Zhu$^{5,1}$ \\
\vspace{4mm}
{\footnotesize{\it
$^1$Institute of Theoretical Physics, Chinese Academy of Sciences, 
           P.O. Box 2735, Beijing, China\\
$^2$CCAST (World Laboratory), P.O. Box 8730, Beijing, China\\
$^3$Department of Physics, Northern Jiaotong University, Beijing,
           China\\ 
$^4$The State Key Lab. of Scientific and Engineering Computing, 
           Chinese Academy of Sciences \\
$^5$National Astronomical Observatories and Beijing Astronomical Observatory,  
           Chinese Academy of Sciences, Beijing, China \\}}
\end{center}
\vspace{6mm}
\begin{quotation}
{\footnotesize
We investigate a two-component model for gravitational lenses, i.e., the
fermion-fermion star as a dark matter self-gravitating system made from two
kinds of fermions with different masses. We calculate the deflection angles
varying from arcseconds to even degrees. There is one Einstein ring. In
particular, we find three radial critical curves for radial magnifications 
and four or five images of a point source. These are different from the
case of the one-component model such as the fermion stars and boson stars. 
This is due to the fermion-fermion star being a two-component concentric 
sphere. Our results suggest that any possible observations of the number of
images more than 3  could imply a polytropic distribution of the mass 
inside the lens in the universe. \\

\noindent PSCA numbers: 98.62.Sb, 95.35.+d, 04.40.-b }
\end{quotation}

\vspace{6mm}
It is suggested that most of the matter in the univers may be dark. Several
types of dark matter distribution, such as local dark matter, galaxy dark
matter, cluster dark matter and background dark matter, exist in the universe
[1]. The dark matter may be consist of bosons or/and fermions [2]. Many
authors have studied the dark matter self-gravitating systems, e.g., fermion
stars [3], boson stars [4], boson-fermion stars [5] and fermion-fermion stars
[6]. The dark matter stars could be formed by ejecting part of the dark
matter, carrying out the excess kinetic energy [7]. This may also be a
mechanism on the formation of such dark matter stars, though a finite
temperature situation is still needed to be studied.  

It is assumed that the only coupling of the dark matter stars to ordinary
matter and radiation is gravitational. So the stars would be transparent
which allows the light to pass through them. General relativity predicts the
deflection of light in gravitational fields, which is the foundation of
compact objects as gravitational lenses. The basic theory of gravitational
lensing was developed by Liebes and others [8]. The first example of
gravitational lensing, twin images QSO 0957+561 A,B separated by 5.7  
arcseconds at the same redshift $z_{s}=1.405$ and mag $\approx$ 17, was 
discovered in 1979 [9]. In 1988 Hewitt et al. [10] observed the first 
Einstein ring MG1131+0456 at redshift $z_{s}=1.13$. Schwarzschild
gravitational lensing in the week gravitational field region is well-known 
[11]. Recently, gravitational lensing effects in strong gravitational field
regions of black holes, neutron stars and boson stars were discussed [12].
However, all the stars have a smooth mass distribution. In the present paper,
we give a two-component model for gravitational lenses. We calculate
gravitational lensing effect for the relativistic fermion-fermion stars (for
the case of week gravitational fields, see [13]). We find that there are
typically four or five images of a point source, being different from the
cases of both the fermion stars and boson stars. This is due to the mass
distribution inside both the boson stars and fermion stars being smooth,
while the fermion-fermion star is a two-component concentric sphere (a
polytropic mass distribution). We also get one tangential critical curve
(Einstein ring) for tangential magnification and three radial critical curves
for radial magnification.  

Consider a two-component model consisting of two types of fermions with
different masses $m_1$ and $m_2$. Each of the two types is assumed to be
a Fermi gas. For the system, Einstein's field equations (in G=c=1 units) read
 $$R_{\mu\nu}-\frac{1}{2}g_{\mu\nu}R =-8\pi\left[ T_{(1)\mu\nu}+T_{(2)
     \mu\nu} \right].                                       \eqno (1)$$
The assumption of the only gravitational interaction inside the system
implies the covariant conservation equations, 
$$T_{(i)~~;\nu}^{~\mu\nu}=0, ~~~~(i=1,2),                   \eqno (2)$$
where 
$$T_{(i)\mu}^{~~~\nu}={\rm diag}(-\rho_{i},~ p_{i},~ p_{i},~p_{i})
                                                             \eqno (3)$$
is the energy--momentum tensor for the {\it i}-th Fermi gas. What is more,
the equations of state (in the parametric form) will be 
$$\rho_{i}=K_{i} ({\rm sh} t_{i} -t_{i}),                \eqno(4)$$ 
$$p_{i}=\frac{1}{3}K_{i} \left({\rm sh}t_{i}-8 {\rm sh}
         \frac{1}{2}t_{i}+3t_{i} \right),                \eqno(5)$$
where
$$ K_i \equiv \pi m_{i}^{4}/4h^{3}.                      \eqno(6)$$
Let
$$K_{2}=\frac{1}{4\pi},                                   \eqno(7)$$
then the unit of length is  
$$\xi =\frac{1}{\pi}\left(\frac{h}{m_{2}c}\right)^{3/2}\frac{c}{(
     m_{2}G)^{1/2}}   =\frac{392}{[m_{2}({\rm eV})]^{2}}{\rm kpc}.  
                                                         \eqno(8)$$
Correspondingly the unit of mass is
  $$\eta =\frac{c^2 a}{G}=\frac{8.18\times 10^{18}}{[m_{2}
        ({\rm eV})]^2}{\rm M_{\odot}}.                    \eqno(9)$$
For the polytropic sphere model, the general metric takes the form
$$ds^2 =e^{\nu}dt^2 -e^{\mu}dr^2 -r^2 (d\theta ^2 +\sin^2 
       \theta d\phi^2)                                  \eqno(10)$$
with ${\nu}$ and ${\mu}$ being functions of the radial distance $r$ from
the center of the star. Finally, we get the basic equations as follows:
$$\frac{dt_{i}}{dr}=- 4\frac{M +4\pi r^3 p}{r(r-2M)} {\rm cth}\frac{1}
  {4}t_{i}, ~~~  ~~~~  \frac{dM_{i}}{dr}=k_{i}r^2 
  ( {\rm sh}t_{i}-t_{i}),                            \eqno(11)$$
and
$$e^{\mu}=(1-2M/r)^{-1},~~~  ~~~ \frac{de^{\nu}/dr}{2e^{\nu}}=
     \frac{M +4\pi r^3 p}{r(r-2M)},                     \eqno(12)$$
where
$$M =M_{1}+M_{2}, ~~~~p=p_{1}+p_{2},                 \eqno(13)$$
with the boundary conditions:
$$t_{i}(r=0)=t_{i0}> 0, ~~~~ M_{i}(r=0)=0,           \eqno(14)$$
and
$$t_{i}(r=R_i )=0                                     \eqno(15)$$
where $R_i $ is the radius of the $i$-th fermion sphere. 

A light ray would be deflected by gravitational field, and then the 
deflection angle of which is given by [14]
$$\hat{\alpha}(r_0)=\Delta\phi (r_0)-\pi,              \eqno(16) $$
with
$$\Delta\phi (r_0)=2\int^{\infty}_{r_0}\frac{e^{\mu/2}}{\sqrt
{\left(r^4/b^2\right)e^{-\nu}-r^2 }} dr                \eqno(17)$$
where $r_{0}$ is the closest distance between the light ray and the center
of the gravitational force, and the impact parameter $b$ is defined by
$$b=r_{0}exp[-\nu (r_{0})/2].                           \eqno(18)$$

For the case of the light ray deflected by a fermion-fermion star, the
equation (17) becomes  
$$\Delta\phi (r_0)=2\int^{R}_{r_0} \frac{e^{\mu/2}}{\sqrt{
  \left(r^4/b^2\right)e^{-\nu}-r^2}}dr
  + 2\int^{\infty}_{R} \frac{1}{\sqrt{\left(r^4/b^2\right) 
  - r^2\left[1-2M(R)/r\right]}}dr.                       \eqno(19) $$
where $R$ is the radius of the star. 

An observer O and a point source S are assumed to be located in an 
asymptotically flat spacetime far away from a fermion-fermion star (sa a lens
L). Let $D_{ol}$ denote the distance from the observer to the center of the
lens, $D_{ls}$ the distance between the lens and the source, and $D_{os}$ the
distance between the observer and the source. An image position is specified
by the angle $\theta$ between OL and the tangent to the null geodesic at the
observer.  $\beta$ stands for the true angular position of the source. The
lens equation may be expressed as [12] 
  $$\sin (\theta -\beta )=\frac{D_{ls}}{D_{os}}\sin\hat{\alpha}. 
                                                       \eqno(20)$$
From the geometry of the lens we have
  $$\sin\theta = \frac{b}{D_{ol}} = 
    \frac{r_0}{D_{ol}}e^{-\nu(r_0)/2}.                  \eqno(21)$$
The magnification of images is given by
  $$\mu =\left(\frac{\sin\beta}{\sin\theta}\frac{d\beta}{d\theta}
         \right)^{-1}.                                  \eqno(22)$$
The tangential and radial critical curves (TCC and RCC, respectively)
follow from the singularities of the tangential
  $$\mu_{t}\equiv \left(\frac{\sin\beta}{\sin\theta}
         \right)^{-1}                                   \eqno(23)$$
and the radial magnification 
  $$\mu_{r}\equiv \left(\frac{d\beta}{d\theta}
         \right)^{-1}.                                  \eqno(24)$$
Then we have from (20)
  $$\mu_{r}^{-1} =\frac{d\beta}{d\theta}=1-\frac{D_{ls}}{D_{os}}
    \frac{\cos\hat{\alpha}}{\cos(\theta -\beta)}\frac{d\hat{\alpha}}
    {dr_0}\frac{dr_0}{d\theta},                         \eqno(25)$$
where $\cos(\theta -\beta)$ can be found from (20), $dr_{0}/d\theta$ can
be got from the derivative of (21) with respect to $r_0$, 
  $$\frac{dr_0}{d\theta}=\frac{2D_{ol}e^{\nu(r_0)/2}\sqrt{1-
    \frac{r_{0}^{2}}{D_{ol}^{2}}e^{-\mu(r_0)}}}{2-r_{0}\frac{d\nu(r_0)
    }{dr_0}},                                            \eqno(26)$$
and $d\hat{\alpha}/dr_{0}$ can be calculated by parametric differential of
(16) and (17) with respect to $r_0$. 

We calculated numerically the angle $\beta$ as a function of the angle
$\theta$, and the tangential and radial magnifications. In the present paper,
we consider the maximal fermion-fermion star. The relevant parameters of the
star are: the mass radio for the two kinds of fermions $m_{2}/m_{1}=5$, the
maximal central gravitational redshift $z_{c}=1.22$, the total mass
$M=1.73\eta$, and the radius $R=16.6\xi$; By using the definitions (9) and
(8) for $\eta$ and $\xi$, the total mass of the maximal star is $M==1.42
\times 10^{19}{\rm M_{\odot}}/[m_{2}({\rm eV})]^2$; The radius is $R=6.51
{\rm Mpc}/[m_{2}({\rm eV})]^{2}$; For example, in case of $m_{2}=10{\rm
eV}$ and then $m_{1}=2{\rm eV}$, $M=1.42\times 10^{17}{\rm M_{\odot}}$ and
$R=65.1\times {\rm kpc}$; for $m_{2}=10{\rm GeV}$ and then $m_{1}=2{\rm GeV}$, 
$M=0.14 M_{\odot}$ and $R=2.0 {\rm km}$. 
 
The angular position $\beta$ of the point source, the tangential
magnification $\mu_t$, and the radial magnification $\mu_r$ are plotted
against the angular position $\theta$ of the image, which being shown in
figures 1--4.

In Fig. 1 (it is assumed that $D_{ls}/D_{os}=1/2$ and $D_{ol}=5\times 10^{5}
\xi$), the continuous curve denotes the plot of the source position angle 
$\beta$ agianst the image position angle $\theta$, and the lines of 
$\beta=$ constants are given by the dashed straight lines. The intersections
between the lines (with $\beta=$constants) and the continuous curve indicate
the angular positions of the images and their numbers from 1 to 5, of which
the ones with $\theta \not= 0$ and $\beta=0$ present the Einstein ring that
corresponds to the tangential critical curve (TCC) coming from the
singularity of $\mu_t$ in Fig. 2. The continuous curve with $\beta >0$ or
$\beta <0$ shows two peaks following from the two-component concentric 
sphere structure of the fermion-fermion star, so that the numbers of the
images may be 4 or 5. The maximal deflection angle $\hat{\alpha}_{max}$
corresponds to the maximum (28 degrees) of $\beta$ at about $\theta=0.25$
arcseconds, and then the maximal reduced deflection angle $|\alpha|=|\theta
-\beta|\simeq 28^{\circ}$. From (20) and  because of $\theta\ll\beta$ and 
$D_{ls}/D_{os}=1/2$, we have the maximal deflection angle $\hat{\alpha}_{max}
\simeq\sin^{-1}(2\sin 28^{\circ})\simeq 70^{\circ}$.    

In Fig. 2 the tangential magnification $\mu_{t}$ is plotted as a function of
the image position $\theta$. The singularity in the magnification $\mu_{t}$
shows the angular position of the Einstein ring (TCC), which is at about
$\theta=542$ arcseconds.  

Figure 3 gives the radial magnification $\mu_{r}$ as a function of $\theta$.
There are three singularities in the magnification $\mu_{r}$ corresponding to
three extreme values (i.e., two maxima and a minimum) on the the continuous
curve of $\beta(\theta)>0$ in Fig. 1, which indicate the angular positions of
the three radial critical curves (RCCs) at about $\theta=0.254, 0.883, 3.811$
arcseconds, respectively. 

Comparing to the one-component models such as the boson stars [12] and the
fermion stars, the results from figures 1 and 3 give us the difference:
There may be four or five images of the point source; At the same time, there
are three radial critical curves. This is due to that the mass distribution
for the fermion-fermion stars are generally not smooth but has two-component
structure. Our results suggest that any possible observations of the number
of images more than 3  could imply a polytropic distribution of the mass
inside the lens in the universe.   

Finally we want to say that the function $\beta (\theta)$ depends upon the
values of $D_{ls}/D_{os}$ and $D_{ol}$, $\beta$ being smaller as $D_{ol}$
larger. In figure 4 we give, as an example, the source position angle $\beta$
as a function of the image position $\theta$ for the case of
$D_{ol}=1.5\times 10^{11}\xi$ and $D_{ls}/D_{os}=1/2$, where the values of
$\beta$ and $\theta$ are of the same order in arcseconds.  

This work was supported partially by The National Natural Science Foundation
of China under Grants Nos. 19745008 and 19835040. 

\begin{center}
{\bf ---------------------------}
\end{center}
\begin{enumerate}

\item[$^*$]\hspace{-2mm}Mailing address; ~~Electronic address:
           yzhang@itp.ac.cn \\ or  zhangyz@sun.ihep.ac.cn 
\item[[1]\hspace{-2mm}] See, e.g., M.S. Turner, in the Proceedings of the
           Second International Workshop on Particle Physics and the Early
           Universe, Asilomar, USA, Nov. 15--20, 1998.
\item[[2]\hspace{-2mm}] L. Roszkowski, in the Proceedings (see [1]);
           hep-ph/9903467. 
\item[[3]\hspace{-1.8mm}] M.A. Markov, Phys. Lett. {\bf 10} 122 (1964);
          J.G. Gao and R. Ruffini, Acta Astrophys. Sin. {\bf 1} (1981) 19;
          C.R. Ching, T.H. Ho and Y.Z. Zhang, Commun. in Theor. Phys.
           (China) {\bf 2} 1145 (1983).                
\item[[4]\hspace{-1.6mm}] M. Colpi, S.L. Shapiro and I. Wasserman, Phys.
           Rev. Lett. {\bf 57} 2485 (1986). 
\item[[5]\hspace{-1.6mm}] A.B. Henriques, A.R. Liddle and R.G. Moorhouse, 
           Phys. Lett. B {\bf 233} (1989) 99; Nucl. Phys. {\bf B337}
           737 (1990).    
\item[[6]\hspace{-1.6mm}] Y.Z. Zhang and K.J. Jin, Phys. Lett. {\bf A128}
           309 (1988); 
           K.J. Jin and Y.Z. Zhang, Phys. Lett. {\bf A142} 79 (1989).  
\item[[7]\hspace{-1.6mm}] E. Seidel and W.M. Suen, Phys. Rev. lett. 
           {\bf 72} 2516 (1994). 
\item[[8]\hspace{-1.6mm}] Jr.S. Leibes, Phys. Rev. {\bf B133} 835 (1964);
           S. Refsdal and J. Surdej, MNRAS {\bf 128} 295 (1964); R.R.
           Bourassa and R. Kantowski, APJ {\bf 195} 13 (1975); for review
           see: R. Narayan and M. Bartelmann, Lectures on gravitational
           lensing, astro-ph/9606001.
\item[[9]\hspace{-1.6mm}] D. Walsh, R.F. Carswell, and R.J. Weymann,
           Nature {\bf 279} 381 (1979).
\item[[10]\hspace{-1.6mm}] J.N. Hewitt, et al., Nature {\bf 333} 537
           (1988).
\item[[11]\hspace{-1.6mm}] P. Schneider, J. Ehlers, and E.E. Falco, 1992,
           Gravitational lenses, Springer Verlag, Berlin.
\item[[12]\hspace{-1.6mm}] K.S. Virbhadra, D. Narasimha, and S.M. Chitre,
           Role of the scalar field in gravitational lensing,
           astro-ph/9801174;  M.P. Dabrowski and F. Schunck, Boson stars
           as gravitational lenses, astro-ph/9807039; K.S. Virbhadra and
           G.F.R. Ellis, Schwarzschild black hole lensing,
           astro-ph/9904193. 
\item[[13]\hspace{-1.6mm}] Z.H. Zhu, K.J. Jin, and Y.Z. Zhang,
           Gravitational lensing effects of fermion-fermion stars: II.
           week field case, submitted. 
\item[[14]\hspace{-1.6mm}] S. Weinberg, 1972, Gravitation and cosmology:
           principles and applications of the general theory of
           relativity, John Wiely \& Sons, NY.
\end{enumerate}    

\newpage
\noindent
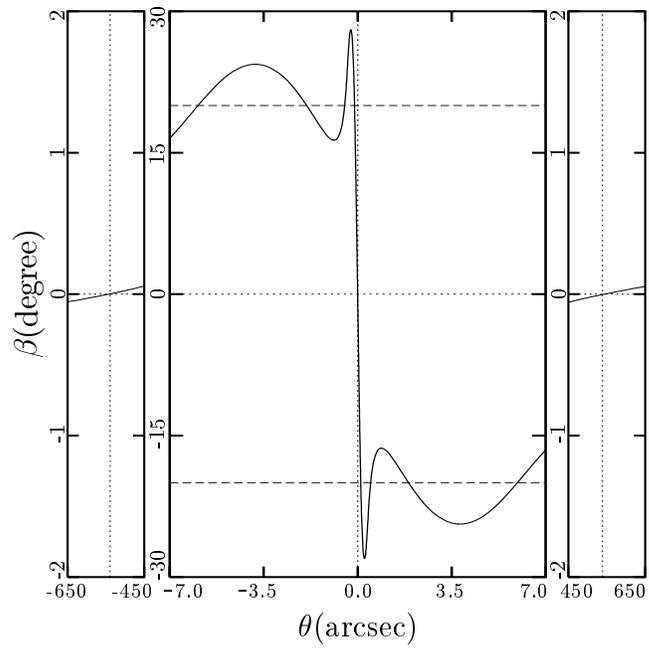
FIG. 1. Gravitational lensing for the maximal fermion-fermion star. The true
angular position $\beta$ of the point source is plotted as a function of the
image position $\theta$ with $(D_{ls}/D_{os})=1/2$ and $D_{ol}=10^{5}\xi$,
which is given by the continuous curve. The values of $\theta$ corresponding
to the interactions between the dashed lines $\beta={\rm constants}$ and the
continuous curve indicate the image positions; It is shown that the number of
the images may be from 1 to 5. The interactions of the straight line $\beta
=0$ with the curve present the Einstein ring at $|\theta |= 542$ arcsecs.
Note that the curve with $\beta>0$ or $\beta<0$ shows two peaks which come
from the two-component structure of the fermion-fermion star.  \\

\noindent
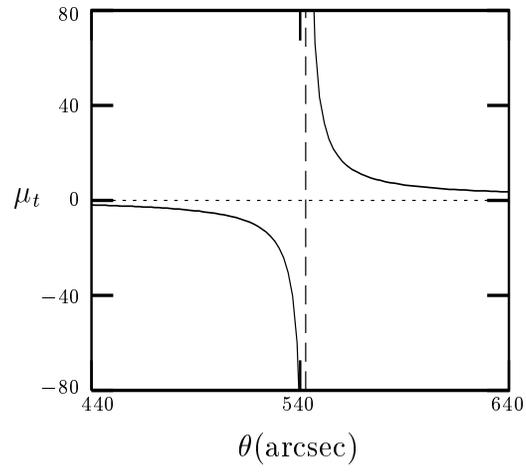
FIG. 2. Corresponding to the curve in Fig. 1, the tangential magnification
$\mu_{t}$ is plotted as a function of the image position $\theta$. \\

\noindent
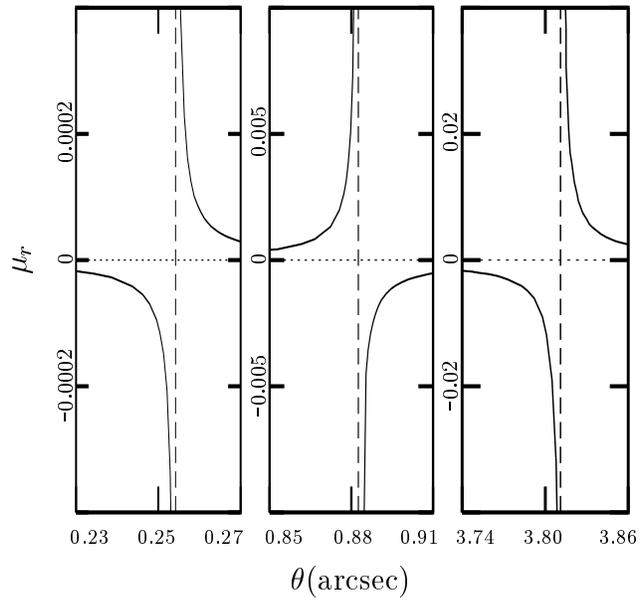
FIG. 3. The radial magnification $\mu_{r}$ as a function of $\theta$. Because
of three extreme values (i.e., two maxima and a minimum) for the the continuous
curve with $\beta(\theta)>0$ in Fig. 1, there are three singularities in the
magnification $\mu_{r}$. \\

\noindent
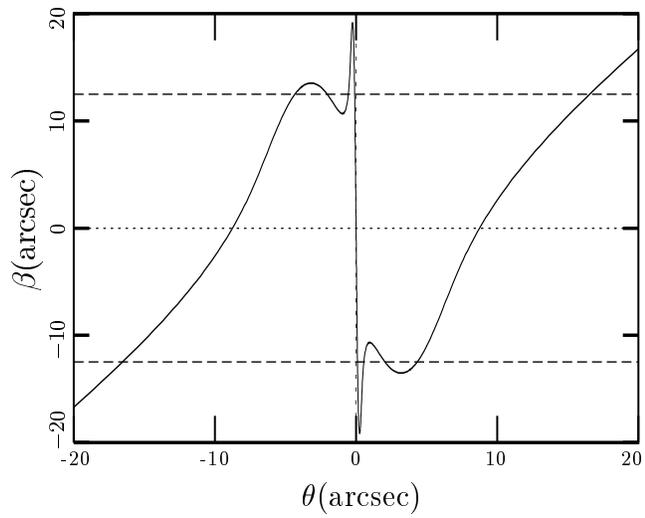
FIG. 4. Gravitational lensing for the maximal fermion-fermion star in the
case of $(D_{ls}/D_{os})=1/2$ and $D_{ol}=1.5 \times 10^{11}\xi$. The angular
positions of the point source and images are all of the same order in
arcseconds.

\end{document}